\documentstyle[12pt]{article}
\newcommand{\beq}{\begin{equation}}
\newcommand{\eeq}{\end{equation}}
\newcommand{\bea}{\begin{eqnarray}}
\newcommand{\eea}{\end{eqnarray}}
\newcommand{\beann}{\begin{eqnarray*}}
\newcommand{\eeann}{\end{eqnarray*}}
\newcommand{\pder}[2]{\frac{\partial #1}{\partial #2}}
\newcommand{\pders}[2]{\frac{\partial^2 #1}{\partial {#2}^2}}

\title{Scattering of second sound waves by quantum vorticity.}

\author{Fernando Lund\thanks{Permanent address: Departamento de F\'\i sica,
Universidad de Chile,
Casilla 487-3, Santiago, Chile} and Victor Steinberg\thanks{Permanent address:
Department of Physics, Weizmann Institute of Science, Rehovot, 76100, Israel}\\
           Laboratoire de Physique, ENS de Lyon \\
            46 Allee d'Italie, 69364 Lyon, France}

\date{}

\begin{document}

\maketitle

\paragraph{Abstract}
 A new method of detection and measurement of quantum vorticity by scattering
second sound off quantized vortices in superfluid Helium
is suggested. Theoretical calculations of
the relative amplitude of the scattered second sound waves  from a single
quantum vortex, a vortex ring, and bulk vorticity are presented.
The relevant estimates show that an experimental verification of the method is
feasible. Moreover, it can even be used for the detection of a single quantum
vortex.

\vspace{4em}
\begin{flushleft}
\end{flushleft}
\newpage

Quantum vorticity in a superfluid, its properties, nucleation and dynamics,
has
been an important and  long standing subject in a quantum fluid physics for
several
decades\cite{donnelly}. The main experimental tool from the very beginning to
investigate
quantum vorticity has been second sound\cite{2sound}. Ever since the first
experiments by Vinen\cite{Vinen} the only method which has been used in this
respect, is the measurement of the variation and anisotropy of second sound
attenuation due to the presence of quantum vorticity\cite{atten}.
This has turned out to be a very useful and powerful tool. However, the
fact that second sound is a radiative hydrodynamic mode suggests alternative
methods to probe quantum vorticity, by analogy with the methods that have been
developed in classical hydrodynamics in
recent years. One of them is to measure the phase difference of a dislocated,
or
splitted, wave front, a phenomenon that is analogous to the Aharonov-Bohm
effect\cite{ab}
This method has been used to probe a bathtub
vortex with water surface waves\cite{surfwave}.
It was also used recently to measure the quantized vorticity of a single vortex
in superfluid helium\cite{single}. Another approach, which is the topic of this
Letter, is
to measure  scattering of sound waves by vorticity\cite{lundrojas}.This method
was successfully
used recently  to measure vorticity in  the vortex array of the Karman street
generated by
flow past a cylinder at low Reynolds number\cite{bcp}.
 In this Letter we present calculations and numerical estimates for the
scattering of second sound waves due to quantum vorticity. Calculations are
valid for any number of vortices both in two and three dimensions, and in three
they may be of arbitrary shape. The estimates convincingly show that the method
discussed is very sensitive and allows for the detection of a single quantum
vortex. This method has the advantages of being practically unperturbative and
fast, so it can be used as a powerful probe to study nucleation of quantum
vortices or vortex rings in superfluid helium.

 A single, straight, isolated quantum vortex consists of a core
of radius $a$, outside of which the flow is potential, with the velocity of the
superfluid component given by
\[
v_s = \nabla \phi
\]
where $\phi = \Gamma \theta /2 \pi $, with $\theta$ the polar angle around the
point vortex, and $\Gamma = h/m$ the quantum of circulation. It is important to
note that this velocity potential is multivalued. At this level the normal flow
is insignificant. Next, a second sound wave is sent onto this vortex. The wave
motion will induce an oscillatory motion of the vortex core which will in turn
radiate second sound. This is the basic mechanism for the scattering of
second sound by a quantum vortex. The scattering of first sound by a quantum
vortex in two dimensions was calculated by Pitaevski\cite{pita1} in a
Born approximation, and by Fetter\cite{fetter} in an exact form. Both used a
hydrodynamic approximation. This issue has
been recently readdressed within the framework of a many body wave
function descrition of elementary exitations in \cite{Ao}.

In order to obtain a first approximation to this problem from a formal point of
view, it is best to fix
one's attention upon a potential velocity field, soon to be identified with the
superfluid velocity, $v_s = \nabla \phi$, whose scalar potential $\phi$  obeys
\beq
\label{eqone}
\pders{\phi}{t} - c_2^2 \nabla^2 \phi = 0
\eeq
with $c_2$ the speed of (second) sound waves, for time scales such that viscous
and thermal conduction effects can be neglected. Vortices are accounted for by
multivalued velocity potentials $\phi$: in two dimensions, a polar angle
centered at ${\vec x}_0$ gives a point vortex in that location; in three
dimensions the solid angle subtended at a given point by a closed curve gives a
vortex loop along that curve. Nonlinear terms can also be
neglected when the second sound frequency is large compared to the time scale
for vortex flow on a length scale comparable to wavelength $\lambda$. Also if
$\lambda \gg a$, the vortex can be considered as a point in two dimensions, or
as a filament in three. The problem of finding the second sound generated by a
vortex filament undergoing prescribed motion is thus reduced to finding the
appropriate multivalued solutions of Eqn. (\ref{eqone}).  The necessary
theoretical framework to handle such problems has been developed by one of
us\cite{lund} in the context of classical hydrodynamics, and the relevant
formulae can be simply translated. The main result to be used is that the
radial
component of the far field velocity, $v_s^r$, generated by a vortex filament
parametrized by $\vec X (\sigma , t)$, where $\sigma$ is a Lagrangean parameter
labeling points along the filament:
\beq
\label{eqtwo}
v_s^r (\vec r , t) = \frac{\Gamma \epsilon_{ijk} {\hat r}_i}{4 \pi c_2^2 |\vec
r|} \pder{}{t} \oint d\sigma  {\dot X}_j X'_k
\eeq
where an overdot means time derivative, a prime means derivative with respect
to
$\sigma$ and the right-hand-side is supposed to be evaluated at the retarded
time $t_{\rm ret} = t - |\vec r|/c_2$. If a plane wave of amplitude ${\vec
v}_s^0$, wave vector ${\vec k}_0$ and angular frequency $\nu_0$ is incident
upon
this filament, it will oscillate with a velocity that is proportional to the
local flow velocity\cite{khalat}:
\beq
\label{eqthree}
{\dot X}_i  = C_{ij} v_{sj}^0 e^{i({\vec k}_0 \cdot \vec X - \nu_0 t)}.
\eeq
and the detrmination of the coefficients $C_{ij}$ will be the subject
of subsequent computation.

Whenever many vortex filaments are present and their mean separation is small
by
comparison with the sound wavelength they can be described in terms of a
continuous vorticity distribution $\vec{\omega} (\vec y , t)$ through the
substitutions
\bea
\Gamma \oint d \sigma X'_j & \longrightarrow & \int d^3 y \omega_j (\vec y , t)
\nonumber \\
\vec x & \longrightarrow & \vec y .
\eea
Using this, and substituting (\ref{eqthree}) into (\ref{eqtwo}) leads to the
following approximate expression for the amplitude ratio $\tilde A$ of wave
scattered into
frequency $\nu$ to incident wave:
\beq
\label{eqfour}
 A \equiv \frac{v_s^r (\vec r , \nu)}{v_s^0} \sim \frac{i e^{i\nu |\vec
r|/c_2}
\nu
\pi^2}{c_2^2 |\vec r|} \epsilon_{jik} {\hat r}_i {\hat k}_{0m} C_{jm}
{\tilde{\omega}}_k (\vec q , \Delta \nu )
,
\eeq
where $\theta$ is the scattering angle, a
tilde denotes Fourier
transform in space and time, $\vec q = (\nu
\hat r /c_2 - {\vec k}_0)$ is the momentum
transfer, $\Delta \nu = \nu - \nu_0$ is the frequency shift and
\[
C_{jm} =(1- \frac{B'}{2}) \delta_{jm} -\frac B2 \epsilon_{jpm} {\hat{\omega}}_p
{}.
\]

Now, there are three important points that must be dealt with in
order to turn the approximate expression (\ref{eqfour}) into an exact
one: The first is that the
oscillatory vortex motion will generate not only second, but also first sound.
The second is that second sound does not obey Eqn. (\ref{eqone}), and the third
is that, in addition to the local interaction between sound and vorticity just
considered, there will be a nonlocal interaction due to the nonlinear coupling
between the sound and vortical motions\cite{nonlin}, as well as additional
effects due to mutual friction.

In order to deal with the first point note that first sound will be radiated at
wavelengths $\sim c_1 /\nu$ and that, as a consequence  the ratio of energy
radiated into first sound waves to that radiated into second sound waves is of
order $ c_2^2 / c_1^2 $. We now choose to consider a situation near the lambda
point, where this ratio is very small and we can neglect the first-to-second
sound scattering.

The second point is that second sound obeys the equation
\beq
\label{eq:secondsound}
\pders{\vec w}{t} - c_2^2 \nabla^2 \vec w = 0 ,
\eeq
where $ \vec w = {\vec v}_n - {\vec v}_s$ is the difference between the normal
and superfluid velocities. However, in a scattering problem we wish to compare
asymptotic states that are plane waves (globally for the incident wave, locally
for the scattered wave), for which the relation
\[
{\vec v}_n = - \frac{\rho_{s0}}{\rho_{n0}} {\vec v}_s
\]
holds. Here ${\rho_{s0}}$ (resp. ${\rho_{n0}}$) is the undisturbed superfluid
(resp. normal) density. Since we have already agreed to be very near the
lambda point,
\beq
v_n \ll v_s
\eeq
and the superfluid velocity obeys the wave equation (\ref{eqone}).
Consequently,
the amplitude $\tilde A$ given by (\ref{eqfour}) really gives the amplitude of
second
sound scattered by quantum vorticity, at least as far as the local interaction
is concerned.

 Since temperature fluctuations are linearly related to superfluid velocity for
a second sound wave, $\tilde A$ will also give the ratio of incident to
scattered
temperature amplitudes. For the situation we have in mind, $c_2 \approx 10^3$
cm/sec. One requirement that must be met in order for (\ref{eqfour}) to be
valid
is that the frequency of the incident wave must be high compared to the time
scales of vortex flow on the length scale of the wavelength of the sound.
Working in the 1-100 KHz range gives wavelengths in the $1-10^{-2}$ cm range,
for which superflow velocities evolve in the $10^{-3}$-$10^{-1}$ s range, so
this requirement is amply met.

 The computation of the scattering of first sound is entirely analogous, it
suffices to replace $c_2$ by $c_1$, the speed of first sound,
in (\ref{eqfour}).
This means that near the lambda point the scattered second sound can easily be
three to four orders of magnitude larger than the scattered first sound.

Dealing with the third point noted above, nonlocal interaction between sound
and vorticity, as well as mutual friction, is harder. The proper way to do this
 is to start with the full
nonlinear two-fluid equations\cite{khalat},
\bea
\label{hvbk}
\pder{\rho}{t} + \nabla \cdot \vec J
& = & 0 \\
\pder{J_i}{t} + \nabla_j \Pi_{ij}  & = & 0 \\
\pder{{\vec V}_s}{t} + {\vec V}_s \cdot \nabla  {\vec V}_s  & = & - \nabla
\mu  + \vec f \\
\pder{(\rho S)}{t} + \nabla \cdot (\rho S {\vec V}_n )
& = & 0
\eea
with
\beann
\vec J & = & \rho_s {\vec V}_s + \rho_n {\vec V}_n \\
\Pi_{ij} & = & p \delta_{ij} + \rho_n {V}_{ni} {V}_{nj} +
\rho_s {V}_{si} {V}_{sj} \\
d\mu & = & \frac{1}{\rho} dp - SdT - \frac{\rho_n}{2\rho} d W^2 \\
\vec f & = & -\frac{B \rho_n}{2 \rho} \vec{\omega}_s \wedge (\hat{\omega}_s
\wedge \vec W) - \frac{B' \rho_n}{2\rho} \vec{\omega}_s \wedge \vec W
\eeann
where $\rho , p, S$ and $T$ are mass density, pressure, entropy per
unit mass and
temperature respectively, ${\vec{\omega}}_s \equiv \nabla \wedge {\vec V}_s$
and $\vec W \equiv {\vec V}_n - {\vec V}_s$.
Thermal as well as viscous diffusivity effects
have been neglected assuming the sound time scale to be small enough. Terms
describing the vortex self-induction have been omitted because they do not
couple to the sound wave. Cubic terms in the entropy equation have been omitted
because they will not contribute in the Born approximation to be considered
shortly.

 The next
step is to describe the flow as the sum of a fast ``sound'' ($\vec v$) and
slow ``vortical'' ($\vec u$) parts:
\[
\vec V = \vec v + \vec u \qquad  v \ll u .
\]
That is, $\vec v$ describes the flow consisting of small deviations of density,
temperature, entropy and pressure away from equilibrium values, around which
the
equations are linearized. These small deviations include both first and second
sound. The
vortical component $\vec u$ is a solution of the equations obtained by
considering constant density, with vortex filaments moving under
their mutual and self-induction, possible external flows and appropriate
boundary conditions. The interaction between the two modes is
obtained by linearizing the equations not around a static value but around the
vortical solution $\vec u$. The result is
\bea
\label{eq:1scatt}
\pders{\rho'}{t} - \nabla^2 p' & = & F_1 \\
\frac{\rho_{n0}}{\rho_{s0} S_0^2} \pders{S'}{t} - \nabla^2 T' & = & F_2
\eea
where $F_1$ and $F_2$ are interaction terms, quadratic in the dynamical
variables, but linear both in $\vec u$ and in the sound variables that are
described by primed quantities, with equilibrium values carrying an index
``0''.
This set of equations has the structure of a wave equation with source terms.
If
the ``sources'' $F_1$ and $F_2$ are discarded, the usual equations describing
sound (both first and second) are obtained. The solution to Eqn.
(\ref{eq:1scatt}) can be written as
\beq
\label{eq:SOUND}
SOUND = SOUND_{\rm inc} + G*F
\eeq
where $SOUND$ is a two-vector describing first and second sound, $G$ is a
two-by-two matrix, the Green's function for the left-hand-side of
(\ref{eq:1scatt}), that can be found with standard Fourier techniques, and $F$
is a two-vector with components $F_1$ and $F_2$. $SOUND_{\rm inc}$ is a
solution
of the homogeneous equation. In an experiment such as we have in mind,
$SOUND_{\rm
inc}$ is a plane, pure second sound wave, the incident wave, and the asymptotic
behaviour of $SOUND$ for large distances gives the scattered wave. It has both
first and second sound components. Near the lambda point, $c_2 \ll c_1$ and
$\rho_{s0} \ll \rho_{n0}$. With these ingredients (\ref{eq:SOUND}) can be
solved
in a first Born approximation that gives for the scattered temperature in the
frequency
domain $T'(\vec x , \nu )_{\rm scatt}$,
\beq
\label{eq:final}
\frac{T'_{\rm scatt}}{T'_0} = \frac{i e^{i\nu |\vec r|/c_2}
\nu
\pi^2}{c_2^2 |\vec r|} \left( H(\theta ) \epsilon_{jmn} {\hat k}_{0m} {\hat
r}_n
{\tilde{\omega}}_j -\frac B2 {\hat r}_i
{\hat k}_{0j} {\tilde \Omega}_{ij} \right)
\eeq
where
\[
H(\theta ) \equiv (1-\frac{B'}{2}) - \frac{1}{2(\cos
\theta -1)} \frac{T_0 S_0}{C
\rho_{n0}} \pder{\rho'_s}{T'} ,
\]
 $C$ is the heat capacity, and
\[
\Omega_{ij} \equiv \frac{\omega_i \omega_j}{\omega} -\omega \delta_{ij} .
\]
Eqn. (\ref{eq:final}) is our main result, for which a first
approximation was provided in Eq. (\ref{eqfour}).

Taking into account the temperature dependence of the superfluid density
$\rho_s$ and of the heat capacity near $T_{\lambda}$\cite{ahlers}, the last
term in the right-hand side of $H(\theta )$ can be estimated to be of order 2-3
at $\tau \equiv (T_{\lambda} -T)/T_{\lambda} \approx 10^{-3}$, and to grow as
$\tau^{0.37}$. This value of $\tau$ is not yet close enough to the
$\lambda$ point for Eqns. (\ref{hvbk}) to loose their validity.

Let us now work out a few simple examples. First, one stationary, straight,
vortex filament will have vorticity given by
\[
\vec{\omega} (\vec y , t) = \Gamma \delta^{(2)} (\vec y) {\hat y}_3
\]
whose Fourier transform is, for $q_3 =0$, which is the only relevant situation
in which the scattering plane is perpendicular to the filament,
\[
\tilde{\omega} (\vec q , \Delta \nu ) = \frac{\Gamma L}{(2 \pi)^3}
\delta(\Delta \nu ) .
\]
where $L$ is the length of the filament or, in the case of a very long
filament,
the portion that is illuminated by the incident sound. Substitution into
(\ref{eqfour}) shows that there is no frequency shift, as was to be expected
for
a static target. Using (\ref{eqfour}) and the frequencies we have quoted gives
a
typical scattered amplitude of $10^{-5}-10^{-7}$, the higher amplitude
corresponding to the higher frequency. This means that with a superconducting
bolometer having a temperature resolution of 10 nK it will be possible to
detect the scattered second sound wave from a single quantum vortex if the
amplitude of the incident wave is higher than 1 mK. These are certainly
experimentally realizable conditions.

Consider next a circular vortex ring moving with velocity $V$. In general, the
vorticity carried by a filamentary vortex $\vec X (\sigma , t)$ is given by
\[
\vec{\omega} (\vec y , t) = \Gamma \oint \delta^{(3)} (\vec x - \vec X (\sigma
,
t)) \pder{\vec X}{\sigma}
\]
and, in the case of a circle, a convenient parametrization is given by
\[
\vec X (\sigma , t) = (R \cos \sigma , R \sin \sigma , Vt ) .
\]
 Take the scattering plane to be the
$y_1 - y_3$ plane. In this case
\bea
q_1 & = &  \frac{\nu}{c_2} \cos \theta - k_0 \nonumber \\
q_2 & = &  0 \nonumber \\
q_3 & = & \frac{\nu}{c_2} \sin \theta
\eea
so that $\tilde{\omega}_2$, the only relevant component, is given by
\[
\tilde{\omega}_2 =
\frac{\Gamma R}{(2 \pi)^3} \delta (\Delta \nu -\frac{V}{c_2} \nu
\sin \theta ) \int_0^{2\pi} d \sigma \cos \sigma e^{-i k_0 R (\cos \theta -1 )
\cos \sigma } .
\]
A Doppler shift in frequency arises from the vortex ring steady
motion at velocity $V$.
For rings whose radii are much smaller that the wavelength, $k_0 R \ll 1$ and
\[
\tilde{\omega}_2 \approx \frac{-i \Gamma k_0 R^2}{8 \pi^2} (\cos \theta -1)
\delta (\Delta \nu -\frac{V}{c_2} \nu
\sin \theta )
\]
so that
\[
A \sim 10^{-7} (k_0 R)^2 ,
\]
which appears to be very small even at 100 KHz and for rings of radius $R \sim
10^{-4}$ cm. Suppose however that we consider $N$ parallel vortex rings
separated by a distance $d$, as might be the case for rings nucleating at a
moving obstacle. In this case the scattering amplitude is enhanced by a factor
$N$ along directions satisfying the Bragg condition
\[
d \lambda \sin \theta = n \qquad n=1,2,\dots
\]

Finally, attenuation techniques have suggested that the quantum turbulent
state\cite{turb}
may be characterized by an anisotropic vortex tangle with a difference in
vorticity parallel and perpendicular to the direction of counterflow given
approximately by\cite{donnelly}
\[
\Delta \omega = \Gamma \alpha \Lambda
\]
where $\alpha$ is the anisotropy coefficient and $\Lambda$ the vortex line
density. Since the anisotropy is of order 1 and the experimental vortex line
density can be as high as 10000, the corresponding scattering amplitude can be
expected to be up to four orders of magnitude higher than that of the single
quantum vortex. Also, in this case it is to be expected that scattering by the
normal fluid vorticity will also be an important effect.

To conclude, we have calculated the scattering of second sound by quantized
vortices in superfluid Helium. Near the lambda point, this is an effect
several orders of magnitude larger than first sound scattering and suggests a
nonintrusive way of probing quantum vorticity.

We wish to thank S. Fauve and Y. Simon for useful discussions, and,
particularly, A. Reisenegger for constructive criticism. The research of F.L.
is
supported in part by Fondecyt Grant 1940429 and CCE Contract CI1*-CT91-0947. V.
S. would like to acknowledge the support of the US-Israeli Binational
Scientific Foundation (BSF)(Grant 90-00412) and The Israel Science Foundation.

\end{document}